\documentclass[twocolumn,showpacs,amsmath,amssymb]{revtex4} % using the revtex package and loading the setup for physical rev. b
\usepackage{graphicx}% Include figure files
\usepackage{dcolumn}% Align table columns on decimal point
\usepackage{bm}% bold math
\usepackage{color}
\usepackage{epstopdf}

\begin{document}

%\preprint{unknown} % don't know yet

\title{%Draft: \\%1.\\
First-principles modeling of temperature and concentration dependent
solubility in the phase separating Fe$_x$Cu$_{1-x}$ alloy system}
\author{D. Reith}
\email{david.reith@univie.ac.at}
\affiliation{
Department of Physical Chemistry, University of Vienna and Center for
Computational Materials Science, Sensengasse 8, A-1090 Vienna, Austria
}
\author{M. St\"ohr}
\affiliation{
Department of Physical Chemistry, University of Vienna and Center for
Computational Materials Science, Sensengasse 8, A-1090 Vienna, Austria
}
\author{T. C. Kerscher}
\affiliation{
Department of Physical Chemistry, University of Vienna and Center for
Computational Materials Science, Sensengasse 8, A-1090 Vienna, Austria
}
\affiliation{
Institute of Advanced Ceramics, Hamburg University of Technology, 
Denickestra\ss e 15, D-21073 Hamburg, Germany
 }
%\author{M. Leitner}
%\email{markus.stoehr@univie.ac.at}
\author{R. Podloucky}
%\email{raimund.podloucky@univie.ac.at}
%\email{email@net}
\affiliation{
Department of Physical Chemistry, University of Vienna and Center for
Computational Materials Science, Sensengasse 8, A-1090 Vienna, Austria
}
\author{S. M\"uller}
\email{stefan.mueller@tuhh.de}
\affiliation{
Institute of Advanced Ceramics, Hamburg University of Technology, 
Denickestra\ss e 15, D-21073 Hamburg, Germany
 }

\date{\today}

\begin{abstract}
We present a novel cluster-expansion (CE) approach for the first-principles modeling of temperature and concentration dependent alloy properties. While the standard CE method includes temperature effects only via the configurational entropy in Monte Carlo simulations, our strategy also covers the first-principles free energies of lattice vibrations. To this end, the effective cluster interactions of the CE have been rendered genuinely temperature dependent, so that they can include the vibrational free energies of the input structures.
%A concept for a first-principles modeling of temperature and concentration
%dependent alloy properties is presented in terms of a Cluster Expansion (CE),
%in which the effective energies are temperature dependent, because vibrational
%free energies for the input structures are included. 
As a model system we use the phase-separating alloy Fe-Cu 
with our focus on the Fe-rich side. There, the solubility is derived from Monte Carlo simulations, whose precision had to be increased by averaging multiple CEs.
We show that including the vibrational free energy 
is absolutely vital for the correct first-principles prediction of Cu solubility in the bcc Fe matrix: The solubility tremendously increases and is now in quantitative agreement with experimental findings.
\end{abstract}

% 71.15.Mb: DFT condensed matter
% 71.15.Nc: binding energy in solids
% 61.72.Qq: Voids (crystal defects)
% 71.20.Eh: Rare earth, electronic structure of ~
% 81.30.Mh: Precipitation in phase transformations
% 63.20.dk: Phonons, first-principle theory

\maketitle
%\input{intro.tex}

%%%%%%%%%%%%%%%%%%%%%%%%%%%%%%%%%%%%%%%%%%%%%%%%%%%%%%%%
%%%%%%%%%%%%%%%%%%%%%%%%%%%%%%%%%%%%%%%%%%%%%%%%%%%%%%%%
% Introduction
%%%%%%%%%%%%%%%%%%%%%%%%%%%%%%%%%%%%%%%%%%%%%%%%%%%%%%%%
%%%%%%%%%%%%%%%%%%%%%%%%%%%%%%%%%%%%%%%%%%%%%%%%%%%%%%%%
First-principles modeling of phase stabilities of alloys is of
scientific and technological importance. A major
progress forward was made by the Cluster Expansion (CE) 
which is based on an Ising-like concept~\cite{Sanchez,ferreira:1989,Muller,Lerch}. The power of CE
consists in modeling concentration dependent properties of coherent
alloy phases based on first-principles input information. For a system 
the energy $E_\text{CE}(\sigma)$ for a particular atomic configuration
$\sigma$ with $N$ atoms is expanded in terms of hierarchical atomic arrangements such as points, pairs, triangles, and higher order objects. Those arrangements are called figures $\text{f}$, and the selected figure set is denoted by $\mathbb{F}$. The CE then reads
\begin{equation}\label{eq:intro1}
E_\text{CE}(\sigma)=N\sum_{\text{f}\in \mathbb{F}}D_\text{f}J_\text{f}{\Pi}_\text{f}(\sigma)~,
\end{equation} 
in which the geometrically determined correlations
${\Pi}_\text{f}(\sigma)$ and the symmetry degeneracy $D_\text{f}$ are
known for the given underlying parental crystal
lattice. 
The unknown effective cluster interaction energies (ECIs) $J_\text{f}$, which
are independent of $\sigma$,
have to be extracted from some suitable input information, such as a
set of density functional theory (DFT) structures, which are denoted by $\sigma\in\text{input}$. For those ordered structures the DFT calculations provide the ground state total energies $E_{0,\text{DFT}}(\sigma)$.  Fitting the CE to the DFT results 
determines the unknown $J_\text{f}$. This is performed by
a least-squares minimization of the residuals~\cite{lu:1991}, which in a simplified formulation \cite{Muller,Lerch} reads
\begin{equation}\label{eq:intro2}
\sum_{\sigma\in\text{input}}| E_\text{CE}(\sigma) -
E_{0,\text{DFT}}(\sigma) |^2 \rightarrow \text{min} ~.
\end{equation} 
The fit is validated by a (leave one out) cross validation score (CVS) \cite{van-de-walle:2002}, which in turn drives a genetic algorithm (GA)  in order to select the optimal figure set $\mathbb{F}$ for the given input \cite{hart:2005,blum:2005,Lerch}.
Additional input is provided until the CE is converged in a self-consistent way.
If the CE in Eq.~\ref{eq:intro1} 
converges reasonably fast 
and the fit in Eq.~\ref{eq:intro2} 
is sufficiently accurate then DFT accuracy can be
carried over to a configuration space much larger than the one defined by the DFT input.
Finally, the combination of CE with Monte Carlo (MC) simulations
%, in which the interactions are
%described by the $J_f$,  
allows a temperature
dependent treatment of phase stabilities and related properties for a very
large number of interacting atoms \cite{kerscher:2011}. 

So far, temperature only
entered via the configurational entropy modeled by 
the MC simulation;  
other temperature dependent contributions were left out, e.g., the important vibrational free energies. 
In the following paragraphs, the present study will include the contributions from lattice vibrations and will demonstrate their strong influence on the phase stability. 
%To include them and
%demonstrate their influence on the phase stability is the aim of the present study. 

Formally, it is obvious that the CE becomes temperature dependent when the ECIs
%$J_\text{f}(T)$ 
become temperature dependent: 
$J_\text{f}\to J_\text{f}(T)$.
This is the result when the ECIs in Eq.~\ref{eq:intro2} are fitted to
temperature dependent input energies.
In the present case those are obtained by summing the temperature dependent vibrational free energy
$F_{\text{vib},\text{DFT}}(\sigma,T)$
to the ground state total energy,
\begin{equation}\label{eq:intro3}
E_\text{DFT}(\sigma,T)
=
E_{0,\text{DFT}}(\sigma)
+
F_{\text{vib},\text{DFT}}(\sigma,T)
\,,
\end{equation}
in which $E_{0,\text{DFT}}(\sigma)$
is the outcome of a standard DFT calculation  strictly valid only at $T=0$~K. 
The label ``DFT'' for $F_{\text{vib},\text{DFT}}(\sigma,T)$
indicates
that it can be derived by the same DFT approach and accuracy as used for
the total energy (see below for details). Other temperature dependent properties
may be included by adding the corresponding temperature dependent terms,
such as the magnetic ordering energy. However, such contributions are not included in the present study and---regarding the magnetic ordering---we assume perfect ferromagnetic ordering in terms of spin polarization.
In order to include these temperature dependent effects,
the CE is rewritten as
\begin{equation}\label{eq:intro4}
E_\text{CE}(\sigma)
\,\to\,
E_\text{CE}(\sigma,T)=N\sum_{\text{f}\in{\mathbb{F}(T)}}D_\text{f}J_\text{f}(T){\Pi}_\text{f}(\sigma)~.
\end{equation} 
Note, that the optimal set of figures has also become temperature dependent: 
$\mathbb{F}\to \mathbb{F}(T)$.

In the following, we will perform and discuss the temperature dependent form of the CE where the additionally included  vibrational free energy   is in general important for the phase stability of alloys and compounds \cite{Garbulsky,Craievich,stohr:2009}. For this purpose the  phase separating binary Fe$_{1-x}$Cu$_x$ alloy system at the Fe-rich side of the phase diagram is considered \cite{LB}. 
For such a 
system the application of CE needs particular care because no ground state line of
ordered compounds exists, i.e. all formation energies are
positive. 
Furthermore, besides the technological interest 
of hardening steel by alloying Fe with Cu, a previous study based on isolated
single-atom and pairwise defects indicated that vibrational free energies are
indeed influential on the solubility of Cu in an Fe matrix \cite{Reith}.
Including vibrational contributions to  CE has been previously discussed 
\cite{VanDeWalle} and applied in very few cases \cite{ozolins:1998b,Yuge}. The actual
procedure, how to include the vibrational free energy is not unique. In
the present work an approach is presented which---in combination with a fast
and accurate procedure for deriving the phonon spectra---can be used in a
convenient way for doing a CE and subsequent MC calculations.

%%%%%%%%%%%%%%%%%%%%%%%%%%%%%%%%%%%%%%%%%%%%%%%%%%%%%%%%
%%%%%%%%%%%%%%%%%%%%%%%%%%%%%%%%%%%%%%%%%%%%%%%%%%%%%%%%
% Methodology
%%%%%%%%%%%%%%%%%%%%%%%%%%%%%%%%%%%%%%%%%%%%%%%%%%%%%%%%
%%%%%%%%%%%%%%%%%%%%%%%%%%%%%%%%%%%%%%%%%%%%%%%%%%%%%%%%

%Cluster Expansion and Monte Carlo
The DFT calculations for the total energies were done by the Vienna \emph{ab
initio} simulation package (VASP) with the pseudopotential construction
according to the projector augmented wave method
\cite{Kresse1,Kresse2,Blochl1}. The exchange-correlation functional was
treated within the  generalized gradient approximation  as parametrized by
Perdew, Burke and Ernzerhof \cite{Perdew}. All calculations were done spin
polarized assuming ferromagnetic ordering of the Fe-atoms.  Very good convergency of total
energies and forces with respect to energy cutoffs and $\vec{k}$-point
integration was ensured. 
%Accurate forces were derived for calculating the
%phonon spectra and vibrational free energies  by a direct force-constant
%method within the harmonic approximation as implemented in the program package
%in the package fPHON \cite{fphon}, which is a modification and generalization of 
%PHON.~\cite{Alfe} 
Accurate forces were derived for calculating the phonon spectra and vibrational free energies by a direct force-constant method within the harmonic approximation as implemented in our program package fPHON, which is based on PHON \cite{Alfe}. 

All the CE and DFT calculations were made for Fe-Cu alloys with a bcc parental
lattice, since the main interest is in the Fe-rich part of the phase diagram
below the ferrite to austenite transition.
For pure Cu, also the fcc ground state total energy was calculated as a
reference.  For the CE the UNiversal CLuster Expansion (UNCLE) program package \cite{Lerch} was applied.
% with the configurational search limited to unit cells containing up to 8 atoms, for which the CE is converged.

%%%%%%%%%%%%%%%%%%%%%%%%%%%%%%%%%%%%%%%%%%%%%%%%%%%%%%%%
%%%%%%%%%%%%%%%%%%%%%%%%%%%%%%%%%%%%%%%%%%%%%%%%%%%%%%%%
% CE
%%%%%%%%%%%%%%%%%%%%%%%%%%%%%%%%%%%%%%%%%%%%%%%%%%%%%%%%
%%%%%%%%%%%%%%%%%%%%%%%%%%%%%%%%%%%%%%%%%%%%%%%%%%%%%%%%

\begin{figure}[btp]
\begin{center}
\includegraphics[width=1\columnwidth]{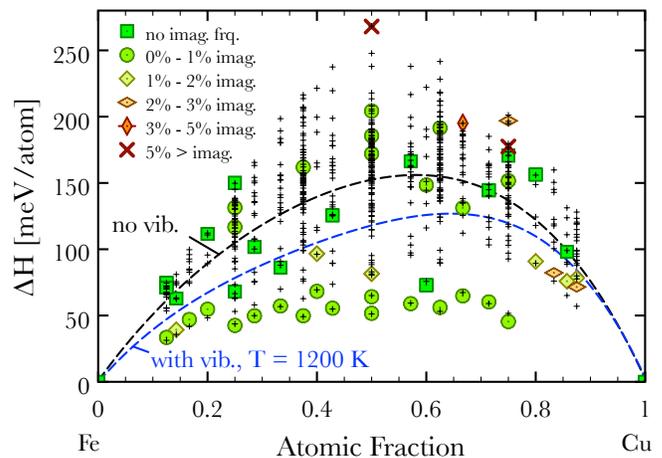}
%\begin{overpic}[width=0.9\textwidth]{file}
%\put(10,-2){(a)} 
%\put(55,-2){(b)} 
%\end{overpic}
\caption{
\label{fig1}
(color online)
Enthalpy of formation
% $\Delta H(x) = E_\text{DFT}(\text{Fe}_{1-x}\text{Cu}_x) - \left[(1-x)E_\text{DFT}(\text{bcc-Fe}) +
%x E_\text{DFT}(\text{bcc-Cu})\right]$, 
derived from $E_\text{CE/DFT}(\sigma,T)$.
% defined by the corresponding DFT total
%energies,  
DFT input values (various symbols) and CE predictions (black crosses) are compared. 
For the phonon calculations of each structure
the percentage of imaginary frequencies is
indicated.  The random mixing energies  are shown for $T=0$~K
(standard CE, black dashed curve) and for $T=1200$~K (CE with vibrational free
energy; blue dashed curve).
}
\end{center}
\end{figure}

%Cluster Expansion
Initially, a standard CE for a bcc parental lattice was made utilizing
only the DFT total energies for $T=0$ K.  The results in  Fig.~\ref{fig1} reveal that no stable binary phase for any composition exists, as
it is expressed by the positive formation energies.  
As expected \cite{LB,Reith}, the configurations with
the lowest formation enthalpies (and the form of the ground-state line)  correspond to phase separating atomic
arrangements, which consist of slabs of pure Cu and Fe.
%Hence, possibly important  Fe-Cu interactions are missing and because of that
%configurations, which are even less favorable \cite{Seko}  have to be
%included.  
In total,  an  input DFT set of 51 configurations was taken into account
resulting in a CVS of $3.7$ meV/atom at $T=0$~K.  
The input set includes the energetically favorable structures as well as configurationally excited states in order to get reliable MC results, cf. Ref.~\citep{Seko}.
In Fig.~\ref{fig1}, we let the CE predict the formation enthalpies 
%\begin{align}
%\Delta H(x) 
%= E_\alpha(\sigma,T) 
%- (1-x)E_\alpha(\text{Fe},T)
%- x E_\alpha(\text{Cu},T)
%\end{align}
of all 631 
%Fe$_{1-x}$Cu$_x$ 
configurations $\sigma$ with unit cells up to 8 atoms large. 
The random mixing energy shown in
Fig.~\ref{fig1} for $T=0$ K (no vibrational free energy included) agrees well with the result of Liu \textit{et al.} \cite{Liu}. 
With increasing temperature (i.e., including
$F_{\text{vib},\text{DFT}}(\sigma,T)$ in the CE) the random energy
is lowered and its maximum shifts to higher Cu concentrations, as shown  in Fig.~\ref{fig1} for $T=1200$~K.

\begin{figure}
\begin{center}
\includegraphics[width = 1\columnwidth]{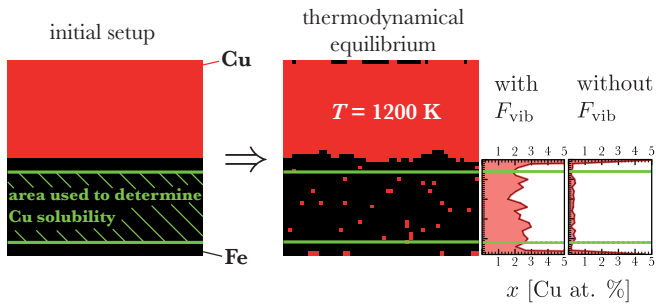}
\end{center}
\caption{\label{fig2} (color online)
Cross section  through the $50\times 50\times 50$ Monte Carlo simulation cell (Fe atoms: black, Cu: red).
The initial setup of pure Cu and Fe blocks as shown on the left panel is
brought into thermodynamical equilibrium for a fixed temperature (right part). 
The volume in the Fe block, in which the
dissolved Cu atoms are counted, is indicated by two
green borders, which are three layers away from the interface.
This ensures that no Cu atom of the Cu slab is erroneously counted as dissolved.
The right most panel demonstrates the concentration of dissolved 
Cu solubility per layer  with and without $F_{\text{vib},{\text{DFT}}}(T)$. 
CE and MC calculations were made for the merged figure set using averaged ECIs (see
text).
}
\end{figure}

Including now  $F_{\text{vib},\text{DFT}}(\sigma,T)$ 
for all 51 structures, Fig.~\ref{fig1} reveals that 
a considerable number of configurations have  phonon spectra with imaginary frequencies, which indicates dynamical instability. Since all configurations are not thermodynamically stable anyway (they have positive formation enthalpies), this is not surprising.
Anharmonic coupling of phonon modes might possibly stabilize some of the phonon
modes \cite{scaild},  but such a task is forbiddingly expensive.  
Therefore, the usual assumption of neglecting non-vibrating modes in the
vibrational free energy is made.

%So far the CE calculations shown in  Fig.~\ref{fig1} were performed for a \textit{single} set $\mathbb{F}(T)$ of figures at $T=0$~K. 
According to Eq.~\ref{eq:intro4}, different temperatures yield different $\mathbb{F}(T)$. However, one finds that the temperature dependence of the solubility is not as smooth  a function of the temperature as expected. This is a direct effect of the GA \cite{hart:2005,Lerch} selecting the figure set $\mathbb{F}(T)$, and it can indeed be likened to that kind of arbitrariness which enters even at a \emph{single} temperature: $n$ different runs of the GA yield  $n$ different $\mathbb{F}_i(T)$. All of them are equally capable to map the input data onto the CE (Eq.~\ref{eq:intro4}) but yield slightly different results in MC simulations. 
For the usual CE applications, this does not pose a problem: the precision needed for MC simulations with respect to concentration is not as strict as needed here for the Cu solubility in Fe ($<1$ at.\%), as we will see later on. In our case we need a strategy which allows us both to find the expected smooth behavior of the solubility and to increase the precision of the prediction. 

We provide the following solution: an averaging procedure either of the results (i.e., the solubilities) or---more physically---of the CEs themselves.
%. For testing purposes,
%10  CEs were made for T=0 K as well as for a
%series of fixed temperatures. It is found that the CVS varies within $\pm
%1\%$, its values starting with 3.7 meV/atom at T=0 K  and increasing to 5.4
%meV/atom at 1200K for the CEs with $F_{\text{vib},\text{DFT}}(T)$.  Also the
%fluctuations of the CVS increase slightly with temperature up to $\pm 1.5\%$
%at 1200K. 
%%The discussed arbitariness is,
%%however, much more noticeable for the solubility at very low concentrations,
%as will be discussed now.
For each temperature, $n=10$ different CEs were 
constructed, with corresponding temperature dependent figure sets $\mathbb{F}_i(T)$ and energies 
$
E_{\text{CE},i}(\sigma,T)
$.
For each CE $i$, a 
separate
MC run was performed, where the simulation took place in a $50\times50\times50$ supercell, starting with the phase separated system by dividing the MC cell into blocks of pure Fe and Cu (see Fig.~\ref{fig2}). 
This setup of fixed reservoirs of Cu and Fe atoms allows for an exchange of atoms between the
two slabs using the  Metropolis algorithm. 
%The standard starting procedure with a system of randomly mixed atoms is very impractical for the dilute limit of a phase separating system, because it needs a huge number of MC steps to reach thermal equilibrium.
Having reached thermal equilibrium at a given temperature, the solubility---i.e., the equilibrium concentration $x_{\text{s}}(\mathbb{F}_i(T))$ of dissolved Cu which depends slightly on the figure set $\mathbb{F}_i(T)$ used---is determined by counting the dissolved Cu atoms in bulk Fe as sketched in Fig.~\ref{fig2}.
% after the equilibrium is reached for a given temperature.
% in which
%the solubility 
%was determined according to Fig.~\ref{fig2}.
For the different CEs, the solubility scatters around the averaged value $\bar{x}_\text{s}(\mathbb{F}_i)=\sum_{i=1}^{n}x_{\text{s}}(\mathbb{F}_i)/n$.
Table \ref{tab2} shows in  the column $\bar{x}_\text{s}(\mathbb{F}_i(T))$  that the fluctuations become sizable at elevated temperatures because  a high precision of the CE is needed to determine the Cu solubility at rather dilute concentrations. Therefore small fluctuations of the CE have a significant impact on the solubility. 
%\cite{japanes paper about accuracy of CE.

%For studying the solubility of Cu MC simulations were performed at given temperatures taking as input
%the temperature dependent ECIs of Eq.~\ref{eq:intro4}, and one has to be
%aware that the figure sets $\mathbb{F}(T)$
%{%Comment--->
%\color{red} 
%are also temperature dependent.
%}%<--Comment
%also become temperature dependent.

\begin{table}
\caption{ Results of 10 temperature dependent CE + MC runs and of one CE +
MC with the merged figure set (see text).
$N_\text{f}$ is the number of figures  in the merged figure set $\bar{\mathbb{F}}(T)$ (see Eq.~\ref{eq:avgCE}). The last two columns show  the Cu solubility as an average value $\bar{x}_\text{s}(\mathbb{F}_i(T))$ of $10$ separate MC runs and as derived from averaged ECIs
 $x_\text{s}(\bar{\mathbb{F}}_i(T))$ (see Eq.~\ref{eq:avgCE}). }
\label{tab2}
\begin{ruledtabular}
\begin{tabular}{l c c c c}
   & T  & $N_\text{f}$  & $\bar{x}_\text{s}(\mathbb{F}_i(T))$ &  $x_\text{s}(\bar{\mathbb{F}}(T))$
\\
   & [K]&   &  Cu at.\%   & Cu at.\%\\%Cu [at.$\%$ ]  \\
\hline
no $F_\text{vib,\text{DFT}}(T)$:   & 1150 & 137&   0.19 $\pm$ 0.04  &  0.18 \\ 
\\
with $F_\text{vib,\text{DFT}}(T)$: &850   & 130&  0.08 $\pm$ 0.03   &  0.06 \\ 
                        &1000  & 125& 0.46 $\pm$ 0.10    &  0.43 \\ 
                        &1150  & 118& 1.58 $\pm$ 0.23    &  1.58 \\
\end{tabular}
\end{ruledtabular}
\end{table}

Instead of running one MC simulation for each of the $n$ CEs, the averaging scheme can also be applied to the CE sums. We note that averaging the results---i.e., determining $\bar{x}_\text{s}(\mathbb{F}_i(T))$---is indeed different from averaging the CEs.
%a scheme can be designed in which only one MC run is needed, namely by 
%averaging the ECIs. 
%{%Comment--->
%\color{red} 
%tk: war hier nicht mal ein Kommentar darueber, warum die beiden Mittelungsmethoden nicht zumselben Ergebnis fuehren muessen.? Das ist IMHO essentiell, das festzustellen. Denn ein Blick auf die Tabelle oder das Phasendiagramm zeigt ja, dass die beiden Methoden quasi das gleich liefern! (Es geht um Abweichungen im 0.05\% Bereich!)
%}%<--Comment
%Then, it has to be 
%%noted
%{%Comment--->
%\color{red} 
%observed
%}%<--Comment
% that
%--because of the genetic algorithm-- each CE has a different number of
%figures. The concept consists in merging all $n=10$ different sets $\mathbb{F}(T)$ into one 
%figure set,
%{%Comment--->
%\color{red} 
The $n$ single CEs (all with their own $\mathbb{F}_i(T)$ and, consequently, their own ECIs) are averaged:
\begin{equation}
\label{eq:avgCE}
\bar{E}_\text{CE}(\sigma,T)
=
\frac{1}{n}\sum_{i=1}^n E_{\text{CE},i}(\sigma,T)
=:
N\sum_{\text{f}\in\bar {\mathbb{F}}(T)} D_\text{f} \bar{J}_\text{f}(T) \Pi_\text{f}(\sigma)
\,.
\end{equation}
On the right-hand side, we introduced the merged figure set 
%$\bar {\mathbb{F}}(T)$ 
%\begin{equation}
%\label{eq:ce1}
$
\bar{\mathbb{F}}(T)=\mathbb{F}_1(T)\cup \dots \cup \mathbb{F}_n(T)
$
%\end{equation} 
with its corresponding temperature dependent averaged ECIs $\bar{J}_\text{f}(T)$.
%It should be noted, that not all figures might occur in all sets, with their
%occurence $n_{occ}\le n$. Now, the ECIs can be averaged
%over the merged figure set by 
%{%Comment--->
%\color{red} 
%tk: das finde ich eigentlich zu kompliziert dargestellt. Wenn es so bleiben soll, wuerde ich auf den mittelteil der Gl verzichten, das verwirrt mich eher als dass es nuetzt.
%}%<--Comment
%\begin{equation}
%\label{eg:ce2}
%\bar{J}_\text{f}(T)
%                =\frac{n_{occ}}{n}\sum_{i=1}^{n} \frac{J_{\text{f},i}(T)}{n_{occ}}
%                =\frac{1}{n}\sum_{i=1}^{n} J_{\text{f},i}(T)~,
%\end{equation}
%As a result only one CE remains now running over all the figures of the merged figure
%set,
%\begin{equation}\label{eq:ce3}
%\bar{E}_\text{CE}(\sigma,T)
%=N\sum_{\text{f}\in \bar{\mathbb{F}}(T)}d_\text{f}\bar{J}_\text{f}(T){\Pi}_\text{f}(\sigma)~.
%\end{equation}  
Obviously, $\bar{\mathbb{F}}(T)$ will
comprise a larger number of figures (more than 100 in the present case, see Table~\ref{tab2}) than any individual CE (about 40 in the present case). 
%On the other hand, the advantage of one merged figure set is that only \textit{one} MC per temperature has to be made.
It should be noted that the value of the ECIs $\bar{J}_\text{f}(T)$ is \textit{not} the result of the CE fitting procedure in Eq.~\ref{eq:intro2} but of the described merging after the fitting.

Table~\ref{tab2} compares the Cu solubilities averaged over $n=10$ MC runs with the result ${x}_\text{s}(\bar{\mathbb{F}}(T))$ of one MC run using the merged figure set $\bar{\mathbb{F}}(T)$ and the averaged ECIs of Eq.~\ref{eq:avgCE}. While both values agree very well  within the error bars of $\bar{x}_\text{s}(\mathbb{F}_i(T))$, it is clear that the two approaches do not yield absolutely the same results, as  already pointed out.

%%%%%%%%%%%%%%%%%%%%%%%%%%%%%%%%%%%%%%%%%%%%%%%%%%%%%%%%
%%%%%%%%%%%%%%%%%%%%%%%%%%%%%%%%%%%%%%%%%%%%%%%%%%%%%%%%
% Phase Diagram
%%%%%%%%%%%%%%%%%%%%%%%%%%%%%%%%%%%%%%%%%%%%%%%%%%%%%%%%
%%%%%%%%%%%%%%%%%%%%%%%%%%%%%%%%%%%%%%%%%%%%%%%%%%%%%%%%
\begin{figure}
\begin{center}
\includegraphics[width = 0.75\columnwidth]{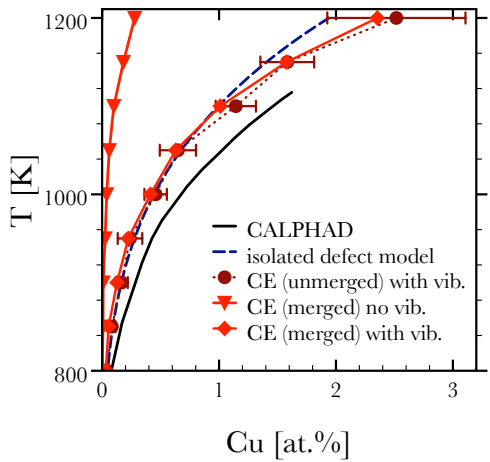}
\end{center}
\caption{\label{fig5} (color online)
Phase boundaries of Fe-rich Fe$_{1-x}$Cu$_x$ alloys.
First-principles results of 10 CEs and one MC  without (triangles, red line)
and with vibrational free  energies, utilizing
temperature dependent ECIs (diamonds, red line) and merged figure sets;
Shown are results averaged over  10 corresponding MCs (red circles, dotted
line) including error bars, and results of a first-principles calculation with
single-atom and pairwise Cu-defects (blue dashed line) ~\cite{Reith}. 
Semi-empirical CALPHAD data~\cite{LB} are indicated as a solid black line .
}
\end{figure}
\begin{figure}
\begin{center}
\includegraphics[width = 1\columnwidth]{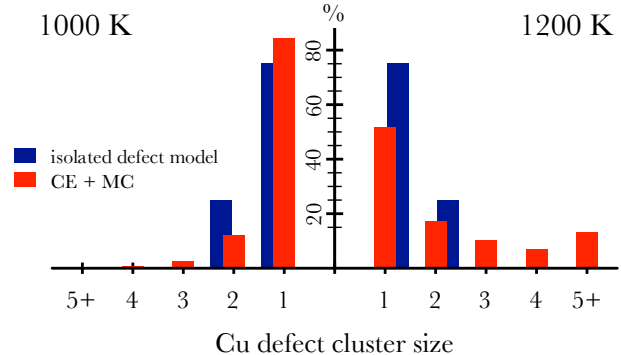}
\end{center}
\caption{\label{fig7} (color online)
Distribution of Cu cluster sizes given as percentage of the total
number of dissolved Cu atoms for the point defect model~\cite{Reith} (blue bars) and the
temperature dependent CE + MC calculation (red bars) with the merged figure set strategy (see text).
}
\end{figure}

Figure~\ref{fig5} presents the phase boundaries at the Fe-rich side. By
comparing the results without and with contributions from the vibrational
free energies  the very striking difference is obvious: without $F_{\text{vib},\text{DFT}}(\sigma,T)$ the
solubility is much too small compared to semi-empirical CALPHAD data \cite{LB}. Obviously,
vibrational entropies are responsible for this effect. 
%The agreement between
%the two averaging procedure---namely, a) averaging the solubilities $\to\bar{x}$ and b) averaging the ECIs $\to\bar{J}_f$---is very good. 
%Procedure
%a) leads to sizable fluctuations due to the genetic algorithm. A
%further advantage of procedure b) is that only one MC has to be done for each
%temperature. On the other hand, the number of figures in procedure b) is 2-3
%times larger than in a).
A comparison of the CE + MC derived phase boundaries to the
 isolated defect model
\cite{Reith} reveals a perfect agreement
at lower temperatures. But at higher temperatures larger defect clusters of Cu atoms
enter the stage, as demonstrated by Fig.~\ref{fig7}.
The CE + MC simulation at 1000$~$K finds most of the
dissolved Cu as  single-atom and pair-wise defects mirroring the
isolated defect model.  Increasing the temperature to
1200$~$K, CE + MC produces a substantial percentage of
larger sized Cu-clusters thus demonstrating the concentration
dependence of this approach and the deficit of the isolated defect model.

%%%%%%%%%%%%%%%%%%%%%%%%%%%%%%%%%%%%%%%%%%%%%%%%%%%%%%%%
%%%%%%%%%%%%%%%%%%%%%%%%%%%%%%%%%%%%%%%%%%%%%%%%%%%%%%%%
% Conclusion
%%%%%%%%%%%%%%%%%%%%%%%%%%%%%%%%%%%%%%%%%%%%%%%%%%%%%%%%
%%%%%%%%%%%%%%%%%%%%%%%%%%%%%%%%%%%%%%%%%%%%%%%%%%%%%%%%

Summarized, we have presented a combination of CE and temperature dependent properties in
terms of vibrational free energies.
With the averaged CEs (Eq.~\ref{eq:avgCE}) a \emph{single} set of ECIs $\bar{J}_\text{f}(T)$ within a merged figure set $\bar{\mathbb{F}}(T)$ has been derived by which one can further study, for example, the growth kinetics of precipitates.
The presented concept for a temperature dependent CE is in principle straightforward and also feasible, in particular if the strategy of the merged figure sets is utilized. Clearly, there is still need for future improvement: in particular, one should aim at reducing the
number of figures in the merged figure set in order to reduce the computational cost of MC simulations. 
In the case of Fe-rich Fe$_{1-x}$Cu$_x$ we have demonstrated that the inclusion of vibrational free energies in the CE+MC simulations is absolutely vital: only then are realistic values obtained for the solubility of Cu in an bcc-Fe matrix, and only then do our results agree with experimental data.
% by making use a kinetic MC scheme.
%No temperature dependent magnetic ordering was included, which is a
%very demanding task. Consequently,  the magnetic phase transition was not considered.
%Nevertheless, 
The main physics behind this surprisingly
large solubility of Cu in Fe is effectively described by a concentration and
temperature dependent and purely first-principles approach which also includes vibrational free energies.

%Acknowledgment
%This work was supported by the austrian science fund FWF under the Grant No. P18480-N19.
%\acknowledgements{
\hfill\break
Work at the University of Vienna was supported by the Austrian Science Fund
(FWF) within the Special Research Program VICOM (Vienna Computational
Materials Laboratory, project no. F4110). Calculations were done on the Vienna Scientific Cluster (VSC) under project no. 70134.
%}

%%%%%%%%%%%%%%%%%%%%%%%%%%%%%%%%%%%%%%%%%%%%%%%%%%%%%%%%
%%%%%%%%%%%%%%%%%%%%%%%%%%%%%%%%%%%%%%%%%%%%%%%%%%%%%%%%
% Bibliography
%%%%%%%%%%%%%%%%%%%%%%%%%%%%%%%%%%%%%%%%%%%%%%%%%%%%%%%%
%%%%%%%%%%%%%%%%%%%%%%%%%%%%%%%%%%%%%%%%%%%%%%%%%%%%%%%%

%\bibliography{CE}

\end{document}